\documentclass[10pt,a4paper,byrevtex,pra,secnumarabic,twocolumn]{revtex4}
\usepackage{SIunits}
\usepackage{graphicx}
\usepackage{psfrag}
\def\kdp{KDP}
\def\kdpfor{\ensuremath{\mathrm{KH}_2\mathrm{PO}_4}}
\def\vpcm{\volt\usk\centi\reciprocal\meter}
\def\jpm{\joule\usk\reciprocal\mole}
\def\ecero{\mbox{\unit{0}{\volt\usk\centi\reciprocal\meter}}}
\def\ecien{\mbox{\unit{100}{\volt\usk\centi\reciprocal\meter}}}
\def\ecuatro{\mbox{\unit{400}{\volt\usk\centi\reciprocal\meter}}}
\def\emil{\mbox{\unit{1000}{\volt\usk\centi\reciprocal\meter}}}
\def\flujo{\phi}
\def\campo{E}

\begin{document}
\title{Influence of the electric field on the latent heat of ferroelectric phase transition in \kdp}
\author{Jose Maria Delgado-Sanchez}\author{Jos{\'e} Mar{\'{\i}}a
  Mart{\'{\i}}n-Olalla}\author{Mar{\'{\i}}a Carmen
  Gallardo}\author{Saturio Ramos}
\affiliation{Departamento de F{\'{\i}}sica de la Materia Condensada. Instituto Mixto de Ciencia de Materiales\\CSIC-Universidad de Sevilla\\Ap Correos 1065, ES-41080 Sevilla, SPAIN}
\author{Marceli Koralewski}
\affiliation{Institute of Physics, Adam Mickiewicz University,
  Umultowska 85, 61-614, Poznan, Poland}
\author{Jaime del Cerro}
\affiliation{Departamento de F{\'{\i}}sica de la Materia Condensada. Instituto Mixto de Ciencia de Materiales.\\CSIC-Universidad de Sevilla\\Ap Correos 1065, ES-41080 Sevilla, SPAIN}
\date{March 30th, 2005}
\preprint{20040426}
\begin{abstract}
The specific heat, heat flux (DTA trace) and dielectric constant of
\kdp\ ferroelectric crystal have been measured simultaneously for
various electric fields with a conduction calorimeter. The specific
heat presents a strong anomaly but these simultaneous measurements
allow us to evaluate the latent heat accurately. Latent heat decreases
with field and the value of critical electric field ---that where
latent heat disappears--- is estimated to be
$\unit{(0.44\pm0.03)}{\kilo\volt\usk\centi\reciprocal\meter}$. Incidentally, we
have measured simultaneously the dielectric permittivity which suggests
that latent heat is developed as domains are growing.
\end{abstract}
\maketitle

\bibliographystyle{apsrev}

\section{Introduction}
\label{sec:intro}

The \kdp{} family is one of the  most extensively
studied\cite{lines-glass-77,xu-91} hydrogen-bonded ferroelectric
crystals. The potassium dihydrogen phosphate \kdpfor{} crystal
exhibits a discontinuous phase transition at $T_0=\unit{121}{\kelvin}$
from a tetragonal paraelectric phase to an orthorhombic ferroelectric
phase. The specific heat anomaly at transition temperature
shows\cite{stephenson-jpc-1944,reese-pr-1967,strukov-pss-1968} a
strong $\lambda$-type anomaly; perhaps, that is why  the transition
was initially considered continuous. However Reese\cite{reese-pr-1967}
showed that the transition is discontinuous by measuring its latent
heat, which was evaluated to be $\unit{46.1}{\jpm}$. It is shown that
discontinuity disappear under the influence of pressure or electric field.

The value of the critical electric field $E_c$ for which the discontinuity disappears is a subject of discussion. Reese et al.\cite{reese-pr-1969} also carried out a measurement of the specific heat with an applied field of $\unit{294}{\vpcm}$ finding a controversial evidence of latent heat. Above $E = \unit{785}{\vpcm}$ they found no evidence of latent heat and suggested $E_c=\unit{300}{\vpcm}$ from the shift of the maximum of the specific heat as function of the electric field.
Measurements of specific heat under electric field higher than
$\unit{360}{\vpcm}$ was carried out by Sandvold and
Fossheim\cite{sandvold-jpc-86} and the authors suggest that Landau theory for continuous transitions with $2-4-6$ potential was appropriate to describe the shape of specific heat curves.

Other works have lead to estimations of the critical field. For instance, Strukov et al.\cite{strukov-spss-72} evaluated $E_c = \unit{124}{\vpcm}$ from electrocaloric experiments. Sidnenko and Gladki\cite{sidnenko-spc-73} found $E_c = \unit{370}{\vpcm}$ while Okada and Sugie\cite{okada-ferro-77} obtained $\unit{160}{\vpcm}$ for $E_c$.  Vallade\cite{vallade-pr-75} deduced a value of $\unit{254}{\vpcm}$ from birefringence measurements. In contrast to these values, Kobayashi et al.\cite{kobayashi-pssb-71} found, by X-ray measurements, $\unit{8000}{\vpcm}$; Eberhard and Horn\cite{eberhard-ssc-75} derived a value of $\unit{6500}{\vpcm}$ from dielectric susceptibility.

In this frame it would be interesting to evaluate the latent heat as a
function of the electric field to determine the value of the critical electric field $E_c$ for which latent heat becomes null.

The difficulty of an accurate determination of $E_c$ from calorimetric measurements in standard equipments ---such as \emph{differential thermal analysis DTA} and \emph{differential scanning calorimeter DSC}---is that these systems really measure changes of enthalpy which has two contributions near the transition: one due to the latent heat and other due to the variation of specific heat with temperature.
In the case of phase transitions near a tricritical point or a
discontinuous ferroelectric phase transition under an electric field
close to the critical field, the specific heat presents a strong
anomaly and latent heat becomes very small. This fact makes difficult
to separate both contributions and to distinguish the temperature interval
where the latent heat is present. This may explain the lack of study
of  the influence of electric field, smaller than $E_c$, on specific heat and latent heat of KDP after Reese\cite{reese-pr-1969}.

Our group has developed a method, named \emph{square modulated
  differential thermal
  analysis} \emph{SMDTA}\cite{jaime-ta-00,javier-ta-01,jaime-ta-03},
based on conduction calorimetry, which is able to measure absolute
values of specific heat and the heat flux exchanged by the sample when
its temperature is changed at a rate as low as
$\unit{0.1}{\kelvin\usk\reciprocal\hour}$. The comparison of the data
allows us to separate the above two contributions to the total
enthalpy and to evaluate the latent heat, in case there were any.

This technique has been successfully applied to the study of the
almost tricritical phase transition of
$\mathrm{KMnF}_3$\cite{jaime-ta-00}, whose latent heat was firstly
measured with this technique. Furthermore, the effect of the
substitution of $\mathrm{Mn}$ by $\mathrm{Ca}$ was also investigated
measuring the latent heat\cite{javier-ta-01}, which showed that the
doping makes the transition become continuous\cite{carmen-mm-00}. The
method of SMDTA has been also applied to show that the phase transition in $\mathrm{CoO}$\cite{javier-jmm-04}, whose character was also controversial, is continuous.

In this paper we have applied this method to study a \kdp{} single crystal. We have measured the specific heat and the heat flux exchanged by the sample in the neighborhood of the ferroelectric phase transition. The measurements have been carried out at four values of $E$: \ecero, \ecien, \ecuatro, \emil. 

Simultaneously to these measurements, the dielectric susceptibility of
the sample has also been measured and we have related its behaviour
around transition temperature with the temperature interval where the
latent heat is produced. Dielectric measurements provide information
about the mechanism of ferroelectric phase transition of KDP crystal\cite{bornarel-ferro-84,nakamura-ferro-92} and the simultaneous measurement of thermal and dielectric properties would be worthy.

\section{Experimental}
\label{sec:experimental}
The measurements were performed in a high resolution conduction calorimeter which has been described previously in details\cite{jaime-jta-88,curro-pt-88,jose-pt-97}. The sensor is formed by two identical heat fluxmeters, each one having 48 chromel-constantan thermocouples connected electrically in series but thermally in parallel. The sample is pressed between both fluxmeters whose signal is measured by a Keithley 182 nanovoltmeter. Two electrodes and two heaters are placed between sample and fluxmeters. 

The sensor is placed inside a calorimeter block which is suspended within two cylindrical radiation shields. The whole assembly is then placed in a hermetic outer case at a high vacuum. The device is then surrounded by a coiled tube and placed in an alcohol bath. Liquid $\mathrm{N}_2$ circulates through the coil and regulates the temperature of the bath with a good thermal stabilization. As a result, it is possible to change smoothly the temperature of the sample (at a rate of about $\unit{0.1}{\kelvin\usk\reciprocal\hour}$) without observing significant temperature fluctuations (always less than $\unit{\microd}{\kelvin}$) in the block temperature.

The specific heat is measured using the method previously described\cite{jaime-ta-03}. The same constant power $W$ is dissipated in both heaters (dissipation branch) for twelve minutes and a steady state characterized by a constant temperature difference between the sample and the calorimeter block is reached. The power is then cut off until a new steady state is reached twelve minutes later (relaxation branch). Then, the power is again switched on and the sequence is continuously repeated while the temperature of the assembly is changed at a low constant rate. That is a long-periodic serial of square thermal pulses is superposed to a heating or cooling ramp. The temperature increase of the sample due to the thermal pulse of twelve minutes (circa $\unit{50\times\millid}{\kelvin}$) is higher than  the temperature variation of the sample produced by the rate of change of temperature in these twelve minutes (circa $\unit{\pm20\times\millid}{\kelvin}$). Hence, the sample is being cooled and heated alternatively during a run.

The integration of the electromotive force  given by the fluxmeter between every pair of steady states allows us to determine sample thermal capacity. Hence, the method is able to determine two data of heat capacity in each cycle. The first one is calculated from the dissipation branch $C_d$, and the second one from the relaxation branch $C_r$. Heat capacity obtained in either branch show a regular behaviour if there is no phase transition or if it is continuous. When a discontinuous phase transition occurs, both data become different showing an anomalous behaviour in the temperature interval where the latent heat is produced as a result of the thermal hysteresis and transition kinetics. That behaviour is an evidence of the discontinuous character of the transition\cite{jaime-ta-03}. Incidentally, we must point out that specific heat data are not reliable when this behaviour is observed.

On the other hand, the DTA trace is continuously measured in a second run without dissipation in the sample and using the same temperature scanning rate used to measure the specific heat. Due to the high number of thermocouples and their good thermal stability of the sample, the equipment works like a very sensitive DTA device. The  electromotive force given by the fluxmeters is proportional to the heat flux, $\flujo_d$ exchanged between sample and calorimeter block.

From the specific heat data obtained in the first running and using a
method previously
described\cite{jaime-ta-00,javier-ta-01,jaime-ta-03}, we calculate the
heat flux $\flujo_c$ which would have been due exclusively to the
behaviour of the thermal capacity of the sample around the transition
temperature. Comparing the measured $\flujo_d$ and the calculated
$\flujo_c$ we deduce that only in the temperature range ($T_f,T_p$)
where both data do not coincide there is an effect from the latent heat. Its value is determined by integrating $\flujo_d/v$, where $v$ is the rate of temperature change, between ($T_f,T_p$) and using the straight line $\flujo_d/v(T_f) - \flujo_d/v(T_p)$ as baseline\cite{javier-ta-01}. The sensitivity of the method is estimated to be better than $\unit{5}{\milli\joule}$.

The single crystal of \kdpfor{} was grown at the Institute of Physics of Poznan
University (Poland). The sample has $\unit{0.3857}{\gram}$ mass, with a thickness of $\unit{2.16}{\milli\meter}$ along the ferroelectric axis and electrodes circular faces with $\unit{78.5}{\milli\meter\squared}$ in surface. The sample was placed out in the calorimeter. Gold electrodes were evaporated of the surface of the sample; those electrodes were connected to a capacitance bridge $ESI-SP 5400$ which has allowed us to measure the dielectric permittivity of the sample, simultaneously to the heat flux, with an imposed external bias field of \ecero, \ecien, \ecuatro\ and \emil.

\section{Results}
\label{sec:results}

\subsection{Calorimetric measurements}
\label{sec:calorimetric}

\begin{figure*}[tb]
  \centering
  \includegraphics[bb=125 423 449 715,angle=270,width=0.9\textwidth]{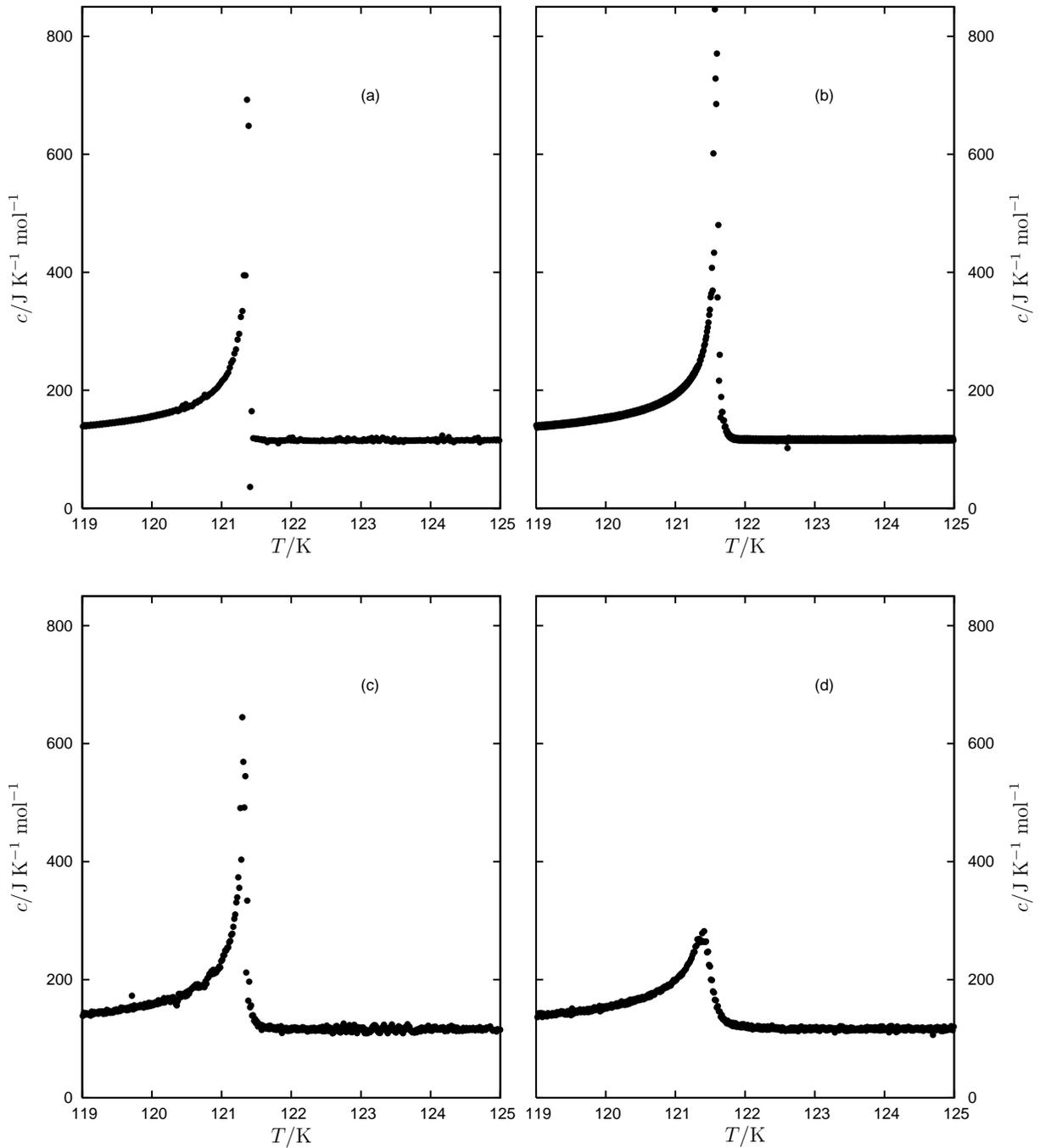}
  \caption{The specific heat of \kdp\ for different applied electric fields in a wide temperature interval. From left to right and top to bottom, (a) \ecero and  (b) \ecien, (c) \ecuatro and  (d) \emil.}
  \label{fig:uno}
\end{figure*}

The temperature dependence of the specific heat $c_p$ of  a sample of
\kdp{} was measured on cooling, on quasiestatic conditions at a
scanning temperature rate of
ca. $v\sim\unit{0.1}{\kelvin\reciprocal\hour}$ using the method
described in Sec.~\ref{sec:experimental} for different applied
electric fields.  In Figure~\ref{fig:uno}, the specific heat data in a
wide temperature interval for different applied electric field  (a) $E
= \ecero$, (b) $E = \ecien$, (c) $E = \ecuatro$ and (d) $E = \emil$ is shown.

These specific heat data show a linear temperature dependence in the
paraelectric phase and for \ecero, \ecien\ and \ecuatro\, a sharp
$\lambda$-type anomaly in a narrow temperature interval is observed. For \emil\, the maximum is more rounded and the transition is smeared. The shape of specific heat curve for high electric field tends to be almost symmetric, such behaviour was also suggested by Reese\cite{reese-pr-1969}.
The maximum value of $c$ initially increases with increasing field but decreases for sufficiently high values. 
As expected, the specific heat tail in the paraelectric  phase increases with field as a consequence  of the coupling of the order parameter to the field.

In Figure 2, we have plotted the specific heat excess obtained in the
dissipation branch ($c_d$) and in the relaxation branch ($c_r$) in a
narrow temperature interval, 1 K, for each electric field. We observed
that for \ecero, \ecien\ and \ecuatro, $c_d$ and $c_r$ data do not coincide around the transition temperature while for \emil{} both series of data almost coincide in the whole range of temperature.
As we have stated above, the temperature variation of the sample due
to the thermal pulses is slightly higher than the variation due to the
temperature ramp. This means that in every period the temperature of
the sample increases and decreases consecutively. Due to thermal
hysteresis, to kinetics of the phase transition etc, the process of
heating and cooling when two phases coexist is different and
consequently data obtained in the dissipation branch and relaxation
branch become different under these conditions. We have reported
previously that the difference is very notorious even in systems near the tricritical point, where the latent heat is very small ( L = \unit{0.13}{\joule\usk\reciprocal\gram} for KMnF$_3$, L = \unit{0.010}{\joule\usk\reciprocal\gram} for KMn$_{0.997}$Ca$_{0.003}$F$_3$)\cite{jaime-ta-03}.

\begin{figure*}[tb]
  \centering
  \includegraphics[bb=125 423 449 715,angle=270,width=0.9\textwidth]{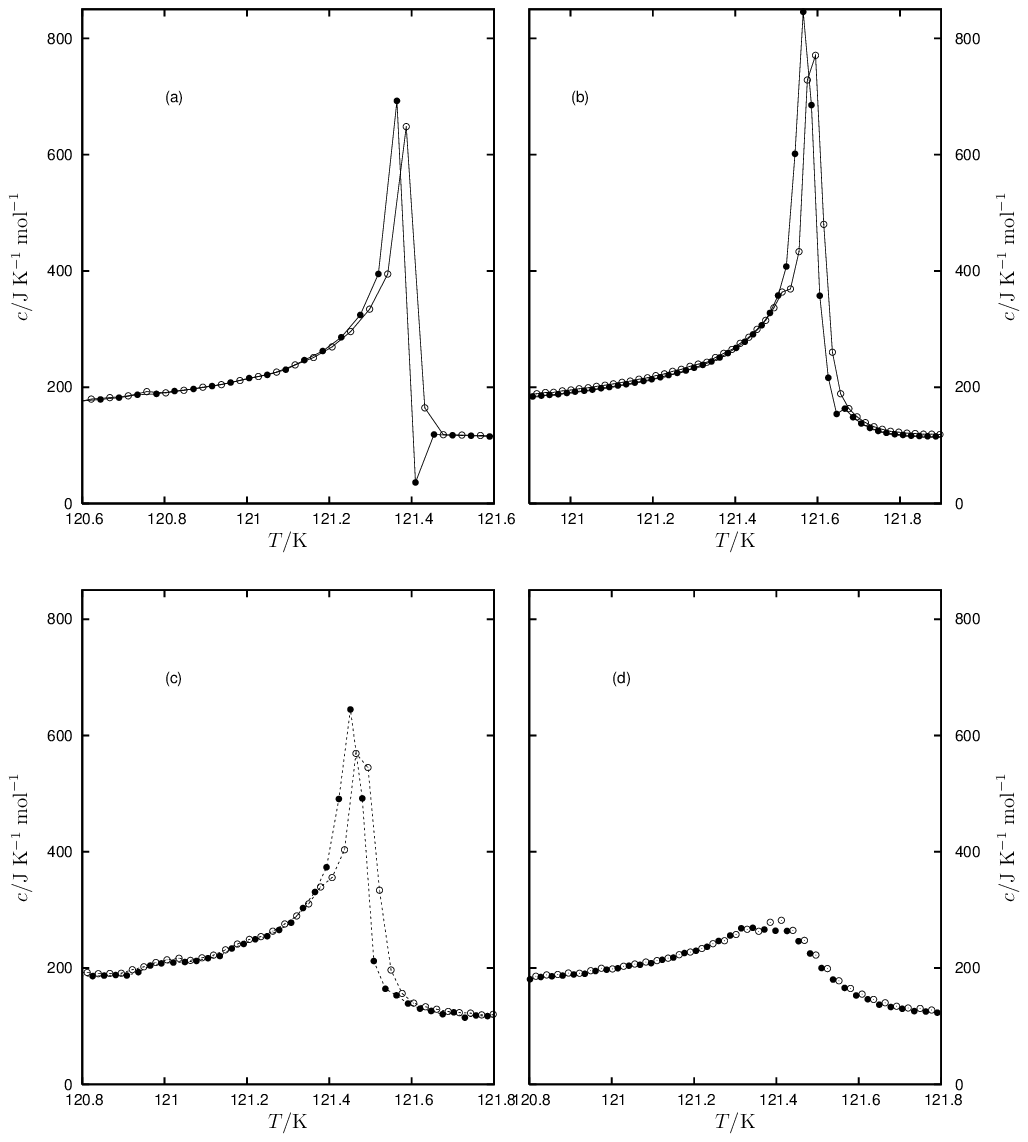}
  \caption{The same as figure~\ref{fig:uno} but in a narrower temperature interval around the transition point. Bold points  stand for the dissipation branch ($c_d$),
open points stand for the relaxation branch ($c_r$).}
  \label{fig:dos}
\end{figure*}

Hence, in the case of KDP, it is clear from Figure~\ref{fig:dos} that
for \ecero, \ecien\ and \ecuatro\ the phase transition is
discontinuous. On the contrary, the similar behaviour of $c_d$ and
$c_r$ for \emil, Figure~\ref{fig:dos}d, indicate that no trace of
latent heat is present, so we can deduce that for \emil\ the phase
transition is continuous.  Hence the critical field lies between \ecuatro\ and \emil.

On the other hand, the baseline respect which the fluxmeter emf is integrated in order to calculate the specific heat is the underline signal due to the temperature ramp imposed on the calorimeter. This signal changes very slowly and it does not affect to the specific heat measurement except when a latent heat effect happens. It produces a non linear variation of the baseline and consequently erroneous data of specific heat may appear. Keeping in mind that the measurements have been carried out in a cooling ramp, in Figure~\ref{fig:dos}a the first $c_r$ at the beginning of the transition which does not coincide with the corresponding $c_d$ data is lower than the regular contribution to the specific heat. In Figure~\ref{fig:dos}b we can also observe such a similar point but this effect is smeared. Finally, in Figure~\ref{fig:dos}c we cannot observe any decrease in $c_r$. Nevertheless, $c_r$ deviates from $c_d$ in a small range of temperature as in the two previous figures. Hence, we can deduce that latent heat decreases with field and should be very small for \ecuatro. 
      
\begin{figure*}[tb]
  \centering
    \includegraphics[bb=150 554 504 747,width=0.9\textwidth]{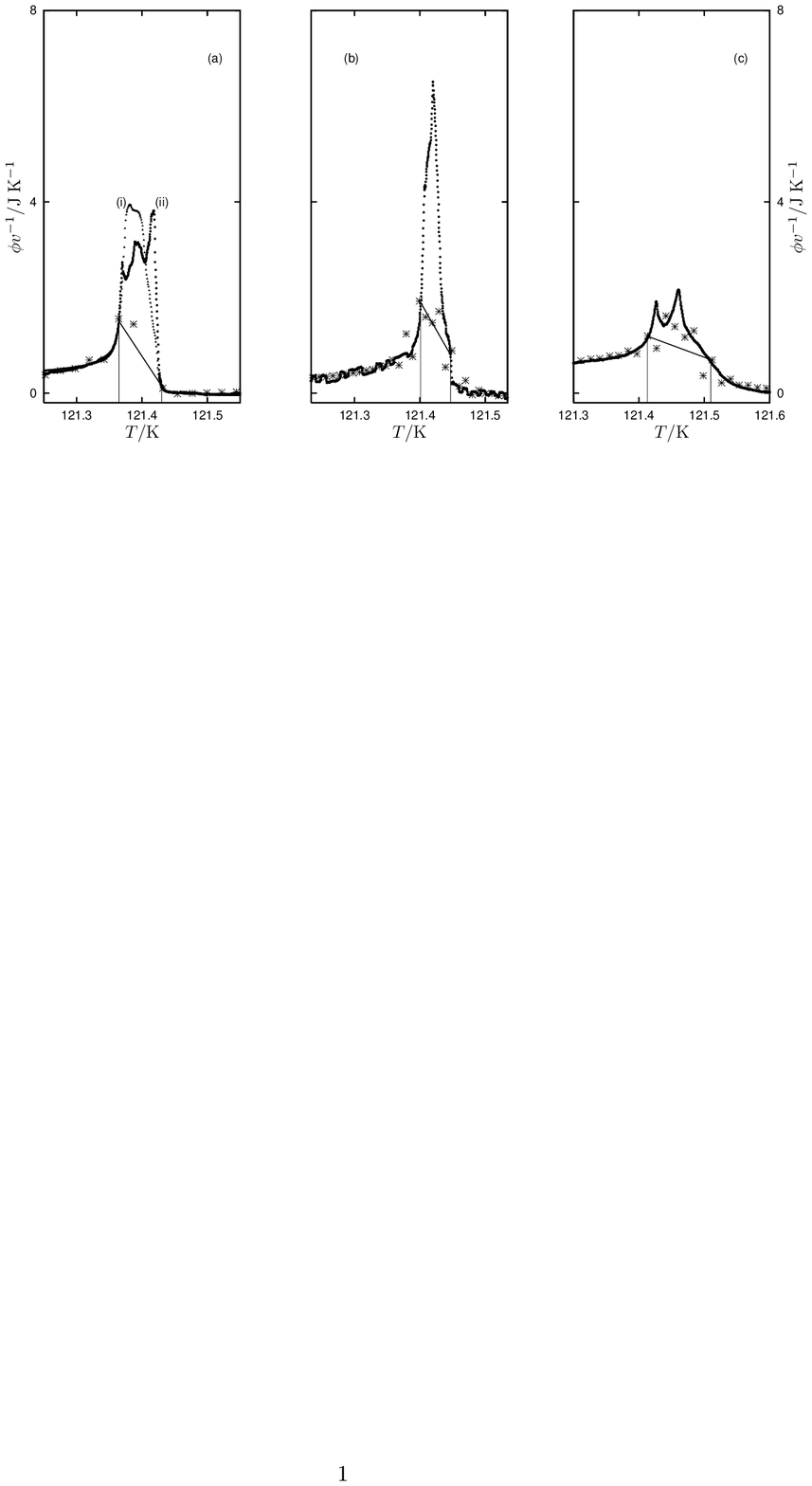}  
  \caption{Heat flux divided by rate of change of temperature in the neighborhood of the \kdp\ ferroelectric phase transition. From left to right, top to bottom, (a) \ecero, (b) \ecien, (c) \ecuatro. Points represent heat flux data given by the fluxmeters ---in (a) two similar runs labelled (i) and (ii) were carried out---. Stars represent the contribution due to specific heat of Figure~\ref{fig:dos}. Straight lines show the baseline used for determination of the latent heat which happens to be the peak area subtended by experimental points and the baseline.}
  \label{fig:tres}
\end{figure*}

 To confirm this suggestion and to calculate the latent heat, we measured the DTA trace in a second run, changing the temperature of the sample at the same constant rate used in the specific heat measurements to make both sets of data comparable. We must point out that the rate of change of temperature is about two orders of magnitude lower than the minimum value achieved in conventional DTA equipments. In Figure~\ref{fig:tres}, we represent the heat flux given by the fluxmeters $\flujo_d/v$ (DTA trace) and the heat flux $\flujo_c/v$ calculated from the specific heat data, using the method previously described, for the fields $\ecero$ (a), $\ecien$ (b), $\ecuatro$ (c).

For $\ecero$ and $\ecien$ $\flujo_d/v$  is higher than $\flujo_c/v$ in a very small temperature interval of about $\unit{0.05}{\kelvin}$, showing the effect of latent heat. For \ecuatro $\flujo_d/v$ is also higher than $\flujo_c/v$, but the difference between them is smaller, indicating a very small latent heat. 

The choosing of the baseline to determine the latent heat value for
$\ecero$ in graph Figure~\ref{fig:tres}a(i) is difficult due to the
very small temperature interval where the latent heat is present and,
consequently, the few data recorded inside that interval. Anyway, by considering as baseline the straight line between the extreme temperatures where $\flujo_d/v$ and $\flujo_c/v$ coincide, we obtain a value of $\unit{43}{\jpm}$. This line is also represented in Figure~\ref{fig:tres}a).
The value obtained for KDP deuterated at $80\%$ using the same set-up
and procedure was \mbox{$\unit{317}{\joule\usk\reciprocal\mole}$}\cite{carmen-jap-97}.

To confirm the validity of that baseline election we again measured the heat flux exchanged by the sample in identical conditions. Those heat flux data are also plotted in the Figure~\ref{fig:tres}a(ii). The figure shows that both heat flux data coincide with $\flujo_c$ (stars) in the same range of temperature thus supporting the baseline used for the determination of the latent heat. In fact, the latent heat for the second run is $\unit{45}{\jpm}$ in good agreement with that of the first one. On the other hand, these values of the latent heat are in agreement with Reese\cite{reese-pr-1969}. 
It is noteworthy that peak area ---i.e. latent heat--- is quite reproducible despite the kinetic effects that showed each experiment. On the contrary, the difference observed if $\phi/v$ during the phase transitions suggests the existence of non-equilibrium kinetic process as those expected for a discontinuous phase transition ---for instance phase front generation--- which, as a general fact, are nonreproducible.

For $\campo=\ecien$ and $\campo=\ecuatro$ we decreased the scanning
rate of temperature when measuring specific heat so as to get a higher
number of data points in the neighborhood of the phase transition. Hence, the determination of the baseline becomes easier as shown in Figure~\ref{fig:tres}. The integration of the peaks gives $\unit{35}{\jpm}$ and $\unit{4.2}{\jpm}$ respectively.

\begin{figure}[tb]
  \centering
    \includegraphics[bb=130 438 500 648,width=0.9\columnwidth]{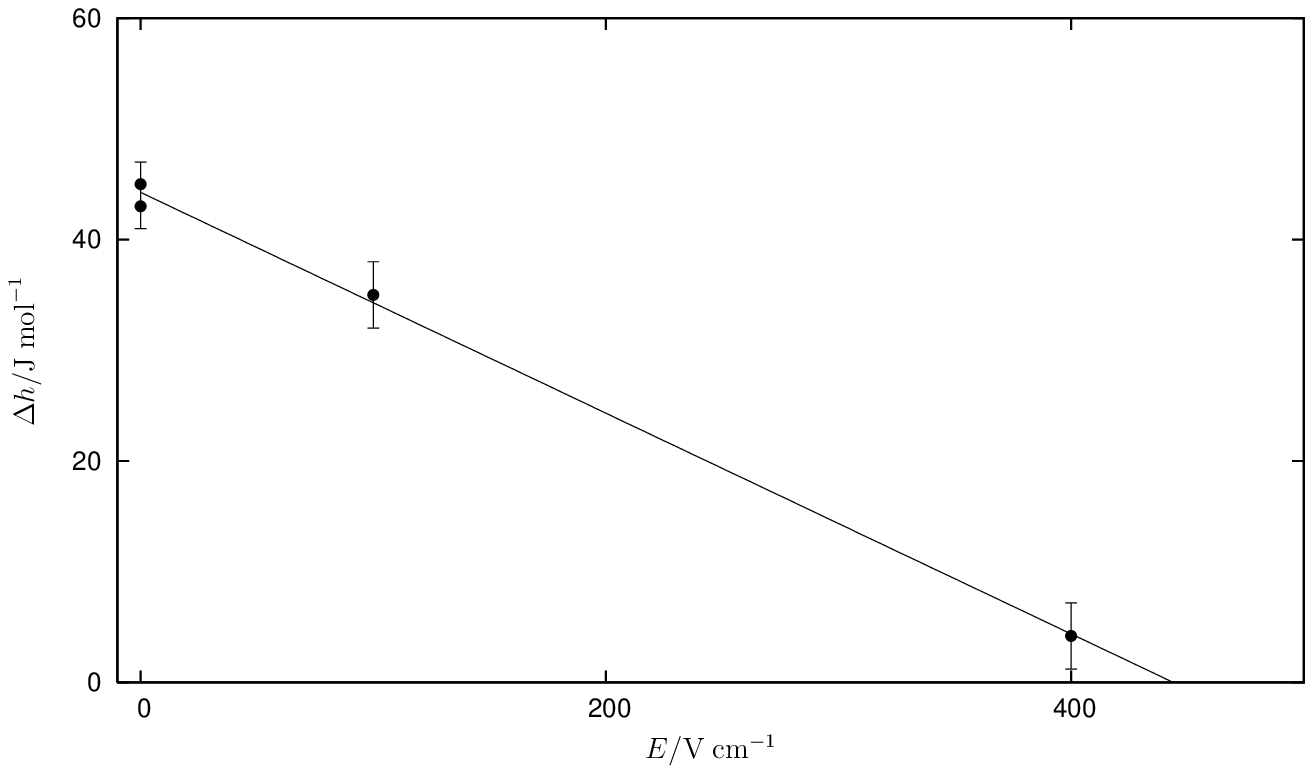}  
  \caption{Plot of the latent heat $\Delta h$ at various electric fields. Critical field is found to be $\unit{(0.44\pm0.03)}{\kilo\volt\usk\centi\reciprocal\meter}$}
  \label{fig:seis}
\end{figure}

The obtained result confirm that latent heat diminishes with electric
field (see Figure~\ref{fig:seis}). At \ecuatro\ latent heat is reduced by and order of magnitude with respect to that of zero field and it is close to the critical field. Assuming a linear behaviour of the latent heat with the field we deduce that the critical field lies on $\unit{(0.44\pm0.3)}{\kilo\volt\usk\centi\reciprocal\meter}$.

\subsection{Dielectric measurements}
\label{sec:dielectric}

In order to relate calorimetric and dielectric behaviour the dielectric permittivity along the ferroelectric axis has been measured simultaneously to the heat flux.

\begin{figure}[tb]
  \centering
  \includegraphics[bb=142 235 482 551,angle=270,width=0.8\columnwidth]{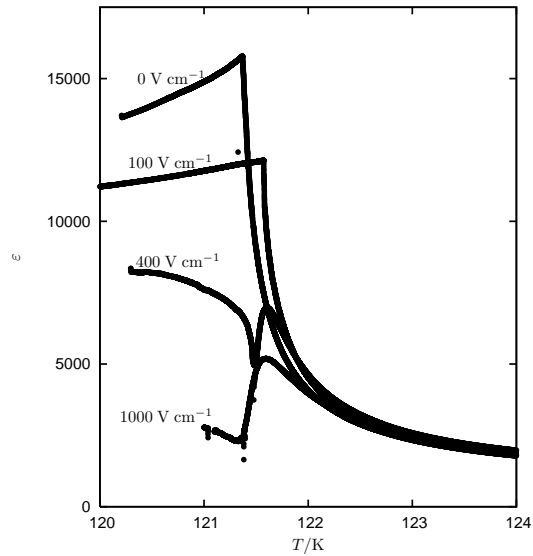}
  \caption{Plot of the dielectric permittivity at various electric
    fields. Permittivity data were measured simultaneously to heat flux data of figure~\ref{fig:tres}.}
  \label{fig:cuatro}
\end{figure}

\begin{figure}[tb]
  \centering
  \includegraphics[bb=142 413 418 587,width=0.9\columnwidth]{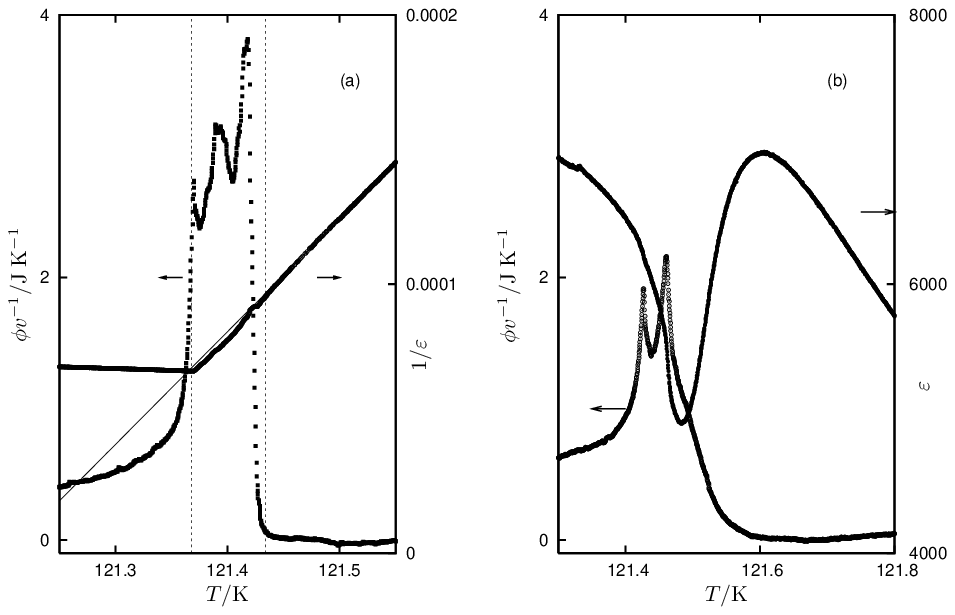}
  \caption{(a) Plot of the inverse of the dielectric permittivity (right axis) and $\phi v^{-1}$ (left axis) for $\campo=\ecero$. (b) Plot of the dielectric permittivity (right axis) and $\phi v^{-1}$ (left axis) $\campo=\ecuatro$. }
  \label{fig:cinco}
\end{figure}

In Figure~\ref{fig:cuatro}  we represent $\varepsilon(T)$ for the
different electric fields. The dielectric permittivity increases in
all cases during the phase transition following the Curie law. In the
ferroelectric phase the dielectric permittivity remains in a plateau,
indicating also the large domain wall contribution to the dielectric
permittivity, but the maximum of permittivity decreases with the
applied electric field. It has been previously reported that the
maximum value of dielectric permittivity is due to the contribution of
domain wall \cite{bornarel-prb-99}; under electric field the sample
becomes close to monodomain state, so number of domain walls decrease
and also does the dielectric permittivity.
Nevertheless the behaviour at transition temperature is different for
$\ecero$ and $\ecien$ ---discontinuous transition--- than for
$\ecuatro$ ---close-to-critical-point transition--- and $\emil$ ---continuous transition---. We will relate in Figure~\ref{fig:cinco} the shape of dielectric permittivity for $\ecero$ and $\ecuatro$.

The heat flux and the
inverse of the dielectric permittivity versus temperature for
$\campo=\ecero$ are represented in Figure~\ref{fig:cinco}(a). Three
regions may be distinguish in this figure: (i) the paraelectric phase where the permittivity follows the Curie law, (ii) the phase transformation interval where the permittivity slightly deviates from the previous behaviour and (iii) the ferroelectric phase where the permittivity shows a plateau. We must point out that the maximum of the permittivity matches with the end of phase transformation.

For $\ecuatro$, $\varepsilon(T)$  shows a different behaviour ---see
Figure~\ref{fig:cinco}(b)---. At the transition temperature it is
present a minimum. This has been observed by
Bornarel\cite{bornarel-ferro-84} in crystals of KDP for higher values
of electric field (about \unit{1}{\kilo\volt\usk\centi\reciprocal\meter}). In that work Bornarel explains this behaviour in terms of some domains arrangements during the phase transition; $\varepsilon(T)$ is described as the sum of the behaviour of $\varepsilon_1(T)$ contribution that correspond to the behaviour of monodomain sample, and the corresponding $\varepsilon_2(T)$ due to the contributions of domains. Moreover $\varepsilon_1(T)$ increases from paraelectric phase and after the maximum at transition temperature decreases to zero in ferroelectric phase; $\varepsilon_2(T)$  increases from zero at transition temperature and remains in a plateau in ferroelectric phase. The sum of both contributions gives the appearance of $\varepsilon(T)$ in figure~\ref{fig:cinco}b.

It may be seen also from the results presented on
Figure~\ref{fig:cinco} that the peak  of the heat flux at transition temperature appear at temperature where the production of domains become dominant. 

\section{Conclusions}

The thermal and dielectric behaviour of \kdp{} crystal near the
temperature of its ferro-paraelectric phase transition has been
simultaneously studied under the influence of electric field. For
\ecero\ the latent heat has been measured twice. Although both
measurements shows a different kinetic, their results are similar and
in good agreement with those obtained by Reese\cite{reese-pr-1969}. Despite the high increase of the specific heat around the transition
temperature, the very small values of the latent heat and the narrow
temperature range ---ca. $\unit{0.1}{\kelvin}$--- where transition is
developed, we have been able to discriminate the contribution of the
latent heat to the total change of enthalpy. Finally, simultaneous
measurement of heat flux and dielectric susceptibility suggest that
the effect of latent heat appears at temperatures range where domains
are growing. Maximum dielectric permittivity was noticed at
temperature where phase transition ends. 

It was established that the latent heat decreases with the field and the critical electric field is estimated to be $\unit{(0.44\pm0.03)}{\kilo\volt\usk\centi\reciprocal\meter}$.

\acknowledgements

We wish to thank Prof. Bornarel for fruitful discussions. This work was supported by Spanish \emph{Ministerio de Ciencia y Tecnolog{\'\i}a} contract number BFM2002-02237.

\end{document}